\documentclass[sigconf]{acmart}

\renewcommand\footnotetextcopyrightpermission[1]{}
\AtBeginDocument{%
  }

\usepackage{multirow}
\usepackage{booktabs}
\usepackage[ruled,linesnumbered]{algorithm2e}
\usepackage[normalem]{ulem}
\usepackage{balance}
\usepackage{threeparttable}
\usepackage{url}

\usepackage{graphicx}
\usepackage{textcomp}
\usepackage{xcolor}
\usepackage{hyperref}
\usepackage{subcaption}

\newcommand{\methodname}{CktEvo}

\begin{document}

\title{\methodname: Repository-Level RTL Code Benchmark for Design Evolution
}

\author{Zhengyuan Shi$^{1,*}$, Jingxin Wang$^{2,*}$, Tairan Cheng$^{1}$, Changran Xu$^{1}$, Weikang Qian$^{2}$ and Qiang Xu$^{1}$}
\email{{zyshi21, trcheng26, crxu25, qxu}@cse.cuhk.edu.hk, {jingxin.wang, qianwk}@sjtu.edu.cn}
\affiliation{%
\institution{\textsuperscript{1}The Chinese University of Hong Kong, \textsuperscript{2} Shanghai Jiao Tong University}
\country{}
}

\begin{abstract}
Register-Transfer Level (RTL) coding is an iterative, repository-scale process in which Power, Performance, and Area (PPA) emerge from interactions across many files and the downstream toolchain. While large language models (LLMs) have recently been applied to hardware design, most efforts focus on generation or debugging from natural-language prompts, where ambiguity and hallucinations necessitate expert review. A separate line of work begins from formal inputs, yet typically optimizes high-level synthesis or isolated modules and remains decoupled from cross-file dependencies. In this work, we present \methodname, a benchmark and reference framework for repo-level RTL evolution. Unlike prior benchmarks consisting of isolated snippets, our benchmark targets complete IP cores where PPA emerges from cross-file dependencies. Our benchmark packages several high-quality Verilog repositories from real-world designs. We formalize the task as: given an initial repository, produce edits that preserve functional behavior while improving PPA. We also provide a closed-loop framework that couples LLM-proposed edits with toolchain feedback to enable cross-file modifications and iterative repair at repository scale. Our experiments demonstrate that the reference framework realizes PPA improvements without any human interactions. \methodname{} establishes a rigorous and executable foundation for studying LLM-assisted RTL optimization that matters for engineering practice: repository-level, function-preserving, and PPA-driven. 
\end{abstract}

\maketitle
\let\thefootnote\relax\footnotetext{* Both authors contributed equally to this research. \\ The corresponding author is Qiang Xu. }


\section{Introduction} \label{Sec:Intro}
RTL coding is a complicated iterative process in practice. Designers continuously build scripts, analyze tool feedback, evolve multi-module codebases, and refine verification assets to hit stringent Power Performance Area (PPA) targets. Motivated by the recent advances in Large Language Models (LLMs), a natural idea is to drop models into this iteration loop to simplify chip design and shorten time-to-market.


Although LLMs have recently been brought to semiconductors~\cite{pan2025survey, he2025large}, gaps remain before they can be reliably deployed in industrial RTL code development. On the one hand, most prior efforts target on generation~\cite{liu2023verilogeval, thakur2024verigen, lu2024rtllm, cui2024origen, huang2024towards, akyash2025rtl, guo2025evoverilog, deng2025scalertl} and debugging~\cite{xu2024meic, wang2025veridebug} from natural-language prompts. Unfortunately, due to the ambiguity in textual prompts and model hallucinations, LLM-generated code often appears plausible yet fails functional, security, or non-functional requirements~\cite{pearce2025asleep}. Therefore, the produced formal hardware description language (HDL) still requires expert review before adoption, limiting truly close-loop design automation. 

On the other hand, a separate track begins from formal inputs, such as C/C++, Python or sub-optimal RTL~\cite{xu2024automated, swaroopa2024evaluating, yu2025spec2rtl} and asks LLMs to improve the code. These workflows support formal verification of generated code to guarantee the functional correctness. Yet these approaches typically remain optimizing the modules represented by high-level synthesize (HLS) code and are decoupled from cross-file dependencies. Crucially, PPA outcomes are emergent properties of the entire repository and toolchain, rather than of isolated files. 

The above limitations highlight a disconnect between existing LLM for hardware coding and industrial demand. In this work, we target on the practice iterative process that begins from an unoptimized golden design and evolves toward PPA constraints. We propose \textit{\methodname}~\footnote{Benchmark is available at: \href{https://github.com/cure-lab/cktevo}{https://github.com/cure-lab/cktevo}}, a high-quality repo-level RTL code benchmark for iterative evolution. Our benchmark supports optimizing multi-module codebases by continuous evolution rather than de novo generation of small-scale designs. \methodname~comprises 11 high-quality RTL designs, each organized as a multi-file Verilog repository, spanning diverse categories such as processors, controllers, high-speed interfaces, and encoders/decoders. 


\begin{figure*}
    \centering
    \includegraphics[width=0.85\linewidth]{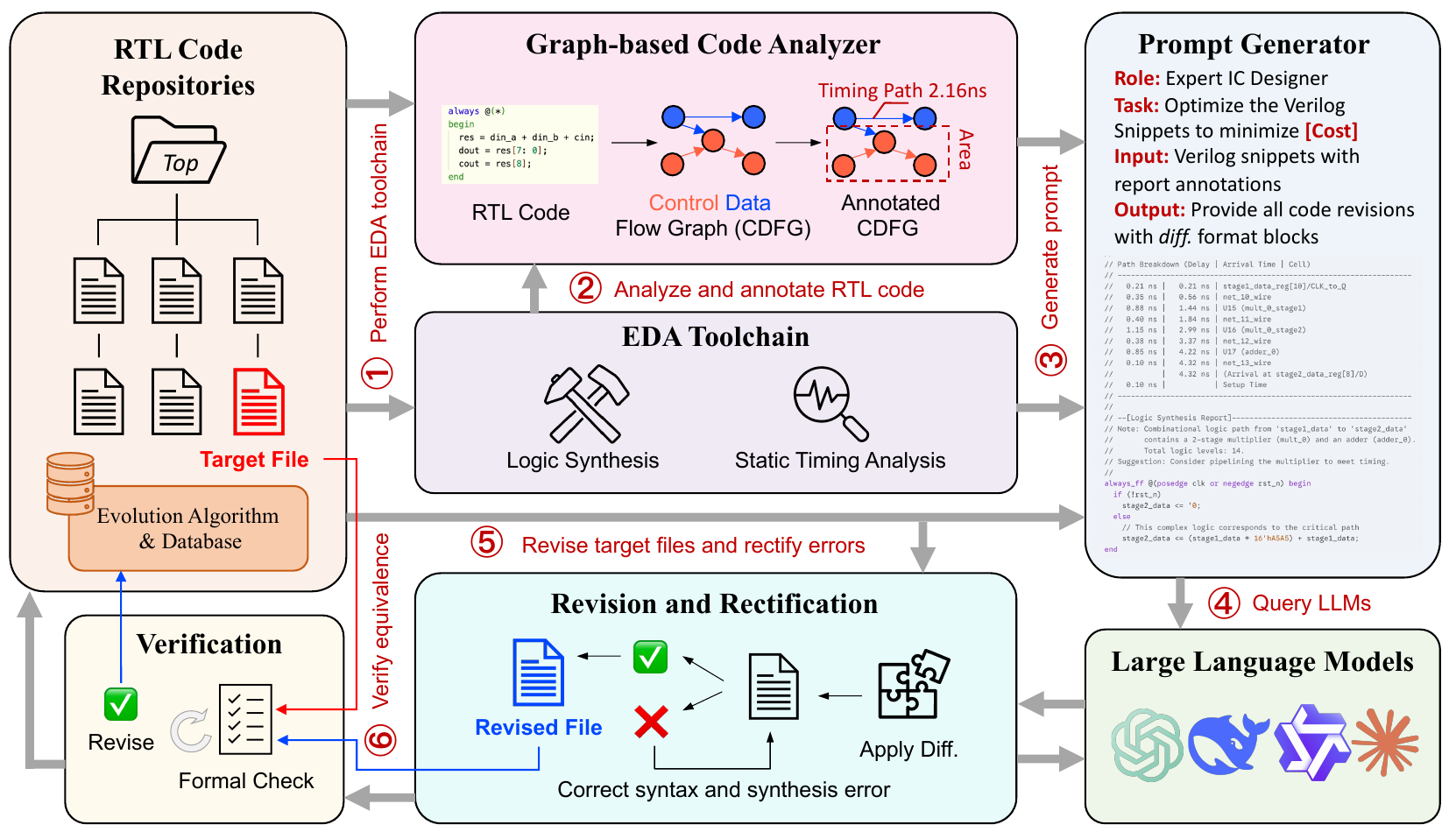}
    \caption{Overview of close-loop evolution framework.} \label{fig:overview}
\end{figure*}

To realize repo-level RTL evolution, we introduce an LLM-driven reference framework. Instead of rewriting all files simultaneously, our framework leverages the insight that system-wide performance is often dictated by bottlenecks within specific modules. Consequently, we integrate the graph-based code analyzer and an efficient prompt generator to pinpoint these bottlenecks reported by EDA tools, and then query LLMs for targeted code improvements (see Figure~\ref{fig:overview}). All LLM-generated revisions are committed only after passing formal verification to guarantee functional equivalence. Our experiments demonstrate that this RTL-level evolution is orthogonal to logic-level optimization, offering additional area and delay reduction. Finally, we highlight key challenges in developing effective evolution agents, positioning \methodname~as a rigorous foundation for future work.


We conclude the contribution of our work as follows: 
\begin{itemize}
    \item We formalize the practical repo-level RTL evolution task. Given the initial design repository, an effective evolution should preserve functional behavior and optimize PPA metrics, tied to end-to-end verification and logic synthesis flows, respectively. 
    \item We release the first repo-level RTL code benchmark for design evolution. The benchmark, named \methodname, comprise 11 high-quality and human-reviewed multi-file Verilog designs spanning various categories. 
    \item We provide a reference framework for closed-loop RTL evolution. This framework supports toolchain feedback, cross-file edits and iterative repair. The experimental results demonstrate that such LLM-based framework can optimize PPA metrics of repo-level RTL code. 
\end{itemize}

\section{Related Work} \label{Sec:Related}
\subsection{LLM for EDA}
With the remarkable generation and textual reasoning abilities of LLMs, the application of LLMs to Electronic Design Automation (EDA) is a burgeoning field of research~\cite{pan2025survey, xu2025large}. A dominant thrust of this research focuses on RTL-level code generation, where the primary goal is to leverage the natural language understanding of LLMs to generate functionally correct Verilog modules from high-level text prompts. To this end, benchmarks like VerilogEval~\cite{liu2023verilogeval}, VeriGen~\cite{thakur2024verigen}, and RTLLM~\cite{lu2024rtllm} have been established to measure the pass-rate of LLM-generated code against a set of module-level problem descriptions. Building on this, many LLM-based RTL code generation approaches and agent frameworks have been proposed in the past several years~\cite{cui2024origen, akyash2025rtl, guo2025evoverilog, deng2025scalertl}. A parallel body of work applies LLMs to the verification and debugging portions of the EDA flow. Moving beyond simple generation, some approaches use LLMs to automatically generate testbenches from design specifications~\cite{huang2024towards, wang2025insights} and repair functional flaws~\cite{fu2023llm4sechw, hassan2025prompt}.
However, these prior efforts are overwhelmingly constrained to small-scale and module-level generation tasks. Until now, generation of an entire circuit design with multiple modules is not yet feasible due to model hallucination. In this work, we focus on benchmarking the LLM for repo-level design optimization instead of generating a single module.

\subsection{LLM for Code Optimization}
In the broader software engineering domain, LLMs are already widely applied to code optimization and refactoring. These "code-to-code" tasks are defined by their premise: they start from a functionally correct codebase and aim to improve non-functional properties while strictly preserving the original behavior. For example, automated program repair iteratively modify code to fix bugs based on test-case feedback~\cite{zhang2024systematic, xia2024automated, bouzenia2024repairagent}. and performance optimization rewrites critical code sections for execution speed~\cite{ishida2024langprop, gong2025language}. Such paradigm of iterative and behavior-preserving software code refinement is now emerging in hardware design. Recent efforts, such as RTLRewriter~\cite{yao2024rtlrewriter} and EvoVerilog~\cite{guo2025evoverilog}, focus on rewriting the given RTL code to optimize the PPA metrics. However, these hardware-focused methods remain limited to module-level and single-file optimization. \methodname~is proposed as the first benchmark to optimize repo-level hardware code. 

\section{Methodology} \label{Sec:Method}
This section details the methodology for our work on repo-level RTL evolution. In Section~\ref{Sec:Method:Benchmark}, we first provide a task definition of the repo-level RTL design evolution: optimizing the delay and area and ensuring the equivalence with the initial design. We then describe the construction of the \methodname~benchmark, including the data source and features of its designs in Section~\ref{Sec:Method:Benchmark}. Finally, we present the architecture of our reference evolution framework, an LLM-based agent designed to perform closed-loop optimization using toolchain feedback in Section~\ref{Sec:Method:Framework}.

\subsection{Task Formulation} \label{Sec:Method:Task}
We formally define repo-level RTL evolution as an iterative, multi-objective optimization problem. Given an initial RTL repository with multiple modules, the task is to find a functionally-equivalent variant of the given ``golden'' design that improves its non-functional hardware metrics. Any new behaviors and properties are also not allowed. 

Let the initial golden design be a repository $\mathcal{D}_0 = (C_0, \Phi)$. Here, $C_0 = \{c_0^1, c_0^2 \ldots, c_{0}^n\}$ is a set of $n$ RTL source files, representing a complete design repository. $\Phi$ is the set of design constraint files (e.g., Synopsys Design Constraints, SDC) used to guide the downstream toolchain. We define the evolution task as an iterative process, generating a sequence of designs ($\mathcal{D}_0, \mathcal{D}_1, ..., \mathcal{D}_T$), where $T$ is the maximum steps. At any step $i$, the current codebase $\mathcal{D}_i$ is evolved based on the previous $\mathcal{D}_{i-1}$ and the constraints $\Phi$ remain constant. 

We prioritize \textbf{area} ($J^{area}$) and \textbf{delay} ($J^{delay}$) as the the primary optimization objectives. While power is a critical design constraint, we do not optimize power as the static power correlates with area and dynamic power is inherently coupled with specific workload vectors (e.g. Switching Activity Interchange Format, SAIF). 

We define two core external functions, representing the EDA toolchains that operate on any given design $\mathcal{D}_i = (C_i, \Phi)$:
\begin{itemize}
    \item \textbf{Verification Function} $\mathcal{V}(C_i, C_0) \rightarrow \{\texttt{True}, \texttt{False}\}$. This function determines if the currently evolved code $C_i$ is still functionally equivalent to the original golden code $C_0$. This function is implemented with logic equivalent checking tools. When the two designs are equivalent, the function returns \texttt{True}, otherwise returns \texttt{False}.
    \item \textbf{Evaluation Function} $\mathcal{E}(\mathcal{D}_i) \rightarrow (J_i, {R}_i)$. This function incorporates the logic synthesis (LS) and static timing analysis (STA) tools to synthesize and analyze the current codebase $C_i$ with design constraint. The output $J_i$ represents quality of current design or cost value of optimization step, which takes both $J^{area}$ and $J^{delay}$ into consideration and supports user customizing. ${R}_i$ is the structured area and timing report, detailing the timing paths, cell areas, module areas and etc. 
\end{itemize}

With the above functions, an agent (e.g., LLM-based agent) with policy function $\mathcal{Q}$ iteratively generates an action transforming the design $\mathcal{Q}(\mathcal{D_i}) \rightarrow \mathcal{D_{i+1}}$ to minimize the cost value $J_{i+1}$. To be specific, the action makes code revision, such as a file-level rewriting or several block-level modifications. An reference agent will be elaborated in Section~\ref{Sec:Method:Framework}.

\subsection{Benchmark Construction} \label{Sec:Method:Benchmark}
We construct a benchmark dataset composed of multiple open-source, multi-file RTL repositories in Verilog. In our benchmark, we provide the initial codebase $C_0$ and a baseline set of synthesis constraints $\Phi$ including timing constraints. 

\begin{table}[!t]
\centering
\caption{Statistics of benchmark designs. } \label{TAB:Benchmark} \vspace{-6pt}
\begin{tabular}{@{}lccc@{}} 
\toprule
Design          & \#Lines & \#Modules & \#Lines/\#Module \\ \midrule
\texttt{audio}           & 979       & 12            & 81.58       \\
\texttt{ethmac}          & 12,285    & 25            & 491.40      \\
\texttt{hsm}             & 2,516     & 16            & 157.25      \\
\texttt{mem\_ctrl}       & 5,894     & 15            & 392.93      \\
\texttt{nn\_engine}        & 4,879     & 15            & 325.27      \\
\texttt{risc}            & 5,209     & 42            & 124.02      \\ 
\texttt{sdc\_ctrl} & 1,550     & 7             & 221.43      \\
\texttt{simple\_cpu}     & 969       & 23            & 42.13       \\
\texttt{spi}             & 611       & 6             & 101.83      \\
\texttt{usb}             & 6,191     & 25            & 247.64      \\
\texttt{vga\_enh}        & 5,095     & 19            & 268.16      \\ \bottomrule
\end{tabular}
\vspace{-10pt}
\end{table}

\paragraph{Data Source} We filter out 11 RTL repositories based on DeepCircuitX~\cite{li2025deepcircuitx} and ForgeEDA~\cite{shi2025forgeeda}, which are the repo-level RTL dataset collected from open-source platforms, such as GitHub and OpenCore~\cite{opencore}. The selection criteria ensures that each design is complete, synthesizable and well-organized. We specifically select designs representing a wide range of design styles and spanning application domains. 

\paragraph{Benchmark} These benchmark designs are summarized in Table~\ref{TAB:Benchmark}, where \#Lines represents the total number of code lines, \#Modules denotes the total count of distinct module definitions within the repository, and \#Lines/\#Module provides the average lines of code per module. Our benchmark spans a diverse range of project scales, from compact designs with hundreds of lines, e.g. a simple Serial Peripheral Interface (SPI) \texttt{spi} for fast evaluation and a basic CPU core \texttt{simple\_cpu}, to large designs. The large design, \texttt{risc}, is a pipelined Reduced Instruction Set Computer (RISC) core, consists of 42 modules, presenting a significant challenge for cross-file reasoning. The benchmark also includes designs from emerging domains, such as a neural network (NN) engine \texttt{nn\_engine} and hardware security module (HSM) \texttt{hsm}. 

\paragraph{Features} The \methodname~benchmark is defined by three features to support the task of repo-level RTL evolution: 
\begin{itemize}
    \item \textbf{Repository-Level and Modular:} Each benchmark instance is a complete, hierarchical repository rather than an isolated module. The codebase $C$ is organized such that each file encapsulates distinct functional units, requiring the agent to manage cross-file dependencies.
    \item \textbf{Toolchain-Agnostic Stability:} All repositories are validated to be synthesizable by both open-source frameworks (Yosys~\cite{wolf2013yosys}) and commercial toolchains. This ensures that the baseline evaluation function $\mathcal{E}(\mathcal{D_i})$ is reproducible and not an artifact of a specific tool version.
    \item \textbf{Production-Grade Baselines:} To avoid strawman comparisons, the selected designs are drawn from established open-source hardware projects and reviewed by our experts. While functionally verified and synthesizable, these designs often reside in a local optimum where traditional EDA synthesis and gate-level optimization tools have saturated. 
\end{itemize}

\subsection{Close-loop Evolution Framework} \label{Sec:Method:Framework}
We introduce an automated closed-loop evolution framework that implements the iterative policy function $\mathcal{Q}$. As illustrated in Figure~\ref{fig:overview}, each iteration begins with the \textbf{EDA Toolchain} (Step \textcircled{1}) running to establish baseline PPA reports by evaluation function $\mathcal{E}(\mathcal{D_i})$. This information is then fed into the \textbf{Graph-based Code Analyzer} (Step \textcircled{2}) and is used by the \textbf{Prompt Generator} (Step \textcircled{3}) to produce prompts. We then query \textbf{Large Language Models} (Step \textcircled{4}) with the context-rich prompt to propose a revision. The code revision is applied, rectified (Step \textcircled{5}), and its functional correctness is verified using formal checking tools (Step \textcircled{6}). The results of this evolution are stored into the evolution database and used to inform the next iteration.

\paragraph{Graph-based Code Analyzer} The raw textual RTL codebase ($C_i$) and output reports ($R_i$) from EDA toolchain from the EDA toolchain are complex and not suitable for direct use by an LLM. The Graph-based Code Analyzer is a pre-processing module that parses the RTL source files $C_i$ into a structured graph-based format, Control Data Flow Graph (CDFG)~\cite{kavi1986formal, namballa2004control}. It then parses the raw tool reports $R_i$ to identify critical timing and area information. For example, the analyzer identifies critical timing paths by analyzing the timing reports from the STA tool and highest-area modules from LS reports. This information is used to annotate the CDFG, creating a direct mapping between high-level synthesis bottlenecks and the specific lines of code and data paths that cause them.

\paragraph{Prompt Generator} The Prompt Generator is responsible for constructing the full context $\mathcal{K}_i$ that is sent to the LLM agent. It acts as a ``prompt engineer'' translating the annotated data from the Code Analyzer, along with relevant RTL code snippets, into a concise context-rich prompt. 

\paragraph{Revision and Rectification} Upon receiving a response from the LLMs, the framework first parses the proposed code modification provided in a \texttt{diff} format and generates a candidate revision. The revised code is not trusted to be immediate valid. It first undergoes a rapid rectification phase, which acts as a low-cost filter and corrects the syntax errors by LLMs. 

\paragraph{Verification} Once a candidate file $C_i^j$ is rectified, the framework proceeds to the final functional verification. This step executes the formal verification function $\mathcal{V}(C_i, C_0)$ to rigorously ensure that the evolved code remains functionally equivalent to the original golden design. If the formal check passes, the revision is accepted as a successful evolution. Otherwise, we automatically make minor revisions with LLMs to revert to the previous code. Since the LLM-driven evolution typically results in localized RTL modifications rather than global rewrites, the formal verification tool only targets on very few modules, significantly reducing the runtime compared to a full verification.

\subsection{Evolution Algorithm and Database}
\begin{algorithm}[!t]
\caption{Dual-Cycle LLM-Guided RTL Evolution}
\label{alg:main}
\KwIn{Initial RTL code $c_0$, max iterations $T$, island count $N_i$}
\KwOut{Optimized RTL code $c^*$}
Initialize archive $\mathcal{A} \gets \{c_0\}$ for each island\;
Initialize best score $J^* \gets J_0$, best code $c^* \gets c_0$\;
\For{$t = 1$ \KwTo $T$}{
    \For{each island $i = 1$ to $N_i$}{
        Sample parent $c_p \sim \text{Uniform}(\mathcal{A}_i)$\;
        Sample inspirations $\{c_j^{ins}\}_{j=1}^k \sim \text{Top-K}(\mathcal{A}_i)$\;
        Extract metrics $\mathbf{r}_p, \{\mathbf{r}_j^{ins}\}_{j=1}^k$ from reports\;
        Construct prompt $\rho \gets \text{Template}(c_p, \mathbf{r}_p, \{c_j^{ins}, \mathbf{r}_j^{ins}\}, H)$\;
        Generate offspring $c_o \gets \text{LLM}(\rho)$\;
        Rectify $c_{o}$ to correct syntax and basic errors\;
        Run LS, STA, and equivalence checking\;
        Compute score $J_o$ and features $\mathcal{F}(c_o) \to b$\;
        Update archive: $\mathcal{A}_i[b] \gets c_o$ if $J_o > J_{\mathcal{A}_i[b])}$\;
        \If{$J_o > J^*$}{
            $J^* \gets J_o$, $c^* \gets c_o$\;
        }
    }
    \If{$t \bmod N_m = 0$}{
        Perform migration between islands\;
    }
}
\Return{$c^*$}\;
\end{algorithm}

As shown in algorithm~\ref{alg:main}, in the close-loop evolution framework, we introduce the evolution algorithm and database to manage the evolutionary search, accumulate success and make further positive evolution. Inspired by AlphaEvolve~\cite{novikov2025alphaevolve}, we uses an LLM as a mutation operator, combined with an island-based population model to enhance search diversity as Algorithm~\ref{alg:main}. 

\begin{table*}[!t]
\centering
\caption{Evolution results of different LLMs with open-source toolchain} \label{Tab:Main} \vspace{-8pt}
    \tabcolsep = 0.01\linewidth
\begin{small}
\begin{tabular}{@{}l|lll|lll|lll|lll@{}}
\toprule
                & \multicolumn{3}{c|}{Baseline}                                     & \multicolumn{3}{c|}{DeepSeek-v3}                                      & \multicolumn{3}{c|}{GPT-4o}                                            & \multicolumn{3}{c}{Qwen3-coder}                                            \\
Design          & Area  & Delay      & ADP               & Area  & Delay       & ADP              & Area  & Delay       & ADP               & Area  & Delay       & ADP               \\ 
          & $(\mu m^2)$  & ($ns$)      & (K)               & $(\mu m^2)$  & ($ns$)       & (K)              & $(\mu m^2)$  & ($ns$)       & (K)               & $(\mu m^2)$  & ($ns$)       & (K)               \\ \midrule
\texttt{audio}           & 29,502.04                    & 1.83       & 53.99     & 27,841.70                    & 1.60       & 44.55             & 29,469.51                    & 1.60       & 47.15             & 28,388.48                    & 1.60       & 45.42             \\
\texttt{ethmac}          & 404,193.90                   & 85.79      & 34,675.80 & 404,031.25                   & 85.64      & 34,601.24         & 402,299.18                   & 84.21      & 33,878.02         & 404,193.90                   & 85.79      & 34,675.80         \\
\texttt{hsm}             & 39,942.06                    & 11.61      & 463.73    & 38,264.20                    & 7.55       & 288.89            & 39,638.02                    & 10.04      & 397.97            & 39,587.97                    & 9.65       & 382.02            \\
\texttt{mem\_ctrl}       & 43,135.12                    & 5.25       & 226.46    & 39,140.04                    & 4.78       & 187.09            & 38,873.53                    & 4.78       & 185.82            & 38,832.24                    & 5.25       & 203.87            \\
\texttt{nn\_engine}        & 1,087,787.02                 & 9.90       & 10,769.09 & 1,086,605.89                 & 8.85       & 9,616.46          & 1,086,605.89                 & 8.85       & 9,616.46          & 1,086,605.89                 & 8.85       & 9,616.46          \\
\texttt{risc}            & 818,703.95                   & 26.91      & 22,031.32 & 816,918.91                   & 24.93      & 20,365.79         & 816,078.93                   & 24.93      & 20,344.85         & 816,744.57                   & 24.93      & 20,361.44         \\ 
\texttt{sdc\_ctrl} & 205,866.19                   & 37.78      & 7,777.62  & 204,209.60                   & 37.78      & 7,715.04          & 206,354.16                   & 30.17      & 6,225.71          & 206,354.16                   & 30.17      & 6,225.71          \\
\texttt{simple\_cpu}     & 104,886.88                   & 8.55       & 896.78    & 104,886.88                   & 8.55       & 896.78            & 91,580.37                    & 8.55       & 783.01            & 90,437.99                    & 8.55       & 773.24            \\
\texttt{spi}             & 4,800.85                     & 0.52       & 2.50      & 4,380.45                     & 0.52       & 2.28              & 4,794.60                     & 0.52       & 2.49              & 4,794.60                     & 0.52       & 2.49              \\
\texttt{usb}             & 187,432.26                   & 2.55       & 477.95    & 187,432.26                   & 2.55       & 477.95            & 186,875.48                   & 2.55       & 476.53            & 187,432.26                   & 2.55       & 477.95            \\
\texttt{vga\_enh}        & 644,244.13                   & 29.05      & 18,715.29 & 636,888.07                   & 27.40      & 17,450.73         & 644,041.44                   & 29.05      & 18,709.40         & 643,980.13                   & 27.40      & 17,645.06         \\ \midrule
Geo.            & 132,931.12                   & 9.01       & \textbf{1,197.55}  & 129,203.02                   & 8.30       & \textbf{1,071.82} & 129,830.02                   & 8.37       & \textbf{1,087.03} & 129,313.40                   & 8.38       & \textbf{1,084.11} \\
Red.            &                              &            &           & 2.80\%                       & 7.92\%     & \textbf{10.50\%}  & 2.33\%                       & 7.06\%     & \textbf{9.23\%}   & 2.72\%                       & 6.94\%     & \textbf{9.47\%}   \\ \bottomrule
\end{tabular}
\end{small}
\end{table*}
\paragraph{Population and Selection}
Following an island-based model to enhance search diversity, we maintain $N_{isl}$ independent populations (islands) $\{\mathcal{I}_1, \ldots, \mathcal{I}_{N_{isl}}\}$, where each island $\mathcal{I}_j$ is a subset of the global archive $\mathcal{A}$, consisting of all designs $\mathcal{D}_i$ equivalent to the golden design. At each iteration $i$, the algorithm selects an island $\mathcal{I}_j$ and samples a parent design $\mathcal{D}_p = (C_p, \Phi)$ from it. Concurrently, $k$ ``inspiration'' designs $\{ (\mathcal{D}_{ins}^{(1)}, {J}_{ins}^{(1)}), \ldots, (\mathcal{D}_{ins}^{(k)}, J_{ins}^{(k)}) \}$ are sampled from the global archive $\mathcal{A}$ to serve as high-quality contextual examples for the agent. 

\paragraph{LLM-Guided Mutation} 
The mutation operator is our LLM-based agent function $\mathcal{G}$. The Prompt Generator (described in Section~\ref{Sec:Method:Framework}) constructs the context for LLMs, which is then formatted into a prompt $\rho_i$. This context is a function of the parent, the inspirations, and the summarized optimization history $H_i$:
\begin{equation} \label{eq:prompt} 
\rho_i \leftarrow \Psi(\mathcal{D}_p, J_p, R_p, { (\mathcal{D}_{ins}^{(j)}, J_{ins}^{(j)}) }_{j=1}^k, H_i) 
\end{equation} 
where $J_p, R_p$ is the tool feedback produced by {evaluation function} $\mathcal{E}(\mathcal{D_p})$. The prompt $\rho_i$ instructs the agent to generate an optimized patch. This patch is applied to the parent code $C_p$ to create an offspring. 

\paragraph{Evaluation Cascade} To balance the high computational cost of the EDA toolchain, we employ a dual-cycle evaluation strategy for the offspring $C_o$. 
\textbf{Fast Evaluation}: The offspring $C_o$ undergoes the rectification process to filter out candidates with clear syntax or basic synthesis errors with a fast LS flow. 
\textbf{Full Evaluation}: This is executed every $N_l$ iterations or when $\mathcal{E}_s$ identifies a promising candidate. In full evaluation, we run the complete evaluation function $\mathcal{E}(\mathcal{D_o})$ and verification function $\mathcal{V}(C_o, C_0)$ to get a trustworthy area and timing reports $R_i$. Only offspring $C_o$ that pass the full evaluation (i.e., $\mathcal{V}(C_o, C_0) = \text{True}$) are eligible for the archive. In our default setting, we define a fitness function as a product of the optimization objectives. Therefore, the evaluation can be considered as reducing the Area Delay Product (ADP) metric. 
\begin{equation} \label{eq:fitness}
    J = J^{area} * J^{delay}
\end{equation}

\paragraph{Archive Update} 
To organize the archive for quality and diversity, we use a MAP-Elites~\cite{mouret2015illuminating} approach. We define a $k$-dimensional behavior descriptor $\mathcal{F}(C) \rightarrow \mathbb{R}^k$, which maps a design's code $C$ to its structural features (e.g., module count, register count, logic depth). This feature space is discretized into bins $\mathcal{B}$. The archive $\mathcal{A}$ stores the best (lowest cost) RTL repository $C_b$ found for each bin $b \in \mathcal{B}$:
\begin{equation} \label{eq:archive}
 \mathcal{A} = \{C_b \mid b \in \mathcal{B}, J_b = \min_{C: \mathcal{F}(C_i) \in b} J_i \}
\end{equation}

For a functionally correct offspring $\mathcal{D}_o$ (with quality $J_o = f(\mathcal{D}_o)$), we compute its bin $b = \mathcal{F}(C_o)$. The archive $\mathcal{A}$ is updated according to the following rule:
\begin{equation} \label{eq:update}
 \mathcal{A}[b] \leftarrow \begin{cases}
 C_o & \text{if } C_o \text{ is new to bin } b \\
     & \text{ or } J_o < J_{\mathcal{A}[b]} \\
 \mathcal{A}[b] & \text{otherwise}
 \end{cases}
\end{equation}



\section{Experiments} \label{Sec:Experiment}
We present a series of experiments to validate the proposed \methodname~benchmark and the effectiveness of our closed-loop evolution framework. 

\subsection{Experiment Settings}
We implement the evolution framework with three representative LLMs: DeepSeek-v3~\cite{liu2024deepseek}, GPT-4o~\cite{achiam2023gpt} and Qwen3-coder~\cite{bai2023qwen}. To ensure a rigorous evaluation, we employ different toolchain setups for the evaluation function $\mathcal{E}$, including the open-source yosys~\cite{wolf2013yosys} (see Section~\ref{Sec:Exp:Yosys}) and commercial EDA toolchain (see Section~\ref{Sec:Exp:DC}). All the benchmark designs are synthesized using Skywater130nm open-source library~\cite{sky130nm}. The verification function $\mathcal{V}$ is implemented using the Synopsys Formality commercial formal verification tool. Due to the page limit, the other LLMs, toolchains and libraries are compatible to \methodname~benchmark but not elaborated. In our experiments, the evolution process is terminated upon reaching either a maximum of $T=50$ iterations or exceeding 4 hours of runtime, which approximates the time required for a human expert to locate and fix a simple design violation. 

\subsection{Results on Open-source Toolchain} \label{Sec:Exp:Yosys}
We evaluate the performance of our closed-loop evolution framework on the CircuitEvolve benchmark. The results obtained by the evaluation function with open-source tools are shown in Table~\ref{Tab:Main}, where ADP is defined as Area $\times$ Delay (scaled by $10^{-3}$). Based on these results, we summarize our findings in three key aspects. 

First, the proposed framework demonstrates universal effectiveness across the entire benchmark suite. Without any manual intervention, the framework successfully reduces the PPA metrics for all design instances. Specifically, the evolution process utilizing the DeepSeek-v3 model achieves an average ADP reduction of 10.50\%. Therefore, iterative evolution can reliably unlock optimization headroom even within functionally complete repositories.

Second, the framework exhibits a significantly stronger impact on delay reduction compared to area reduction. As shown in the DeepSeek-v3 results, the geometric mean reduction for delay is 7.92\%, nearly three times the area reduction of 2.80\%. Notably, the \texttt{hsm} design shows a substantial 34.97\% improvement in timing. We attribute this to the efficacy of graph-based code analyzer. Without the analyzer, the LLM is not able to receive precise instructions to rewrite specific code snippets. 

Third, while generally successful, the magnitude of improvement varies significantly depending on design characteristics. We observe consistent and substantial ADP reductions in control-intensive designs, such as controllers (\texttt{mem\_ctrl}, \texttt{sdc\_ctrl}) and processors (\texttt{simple\_cpu}, \texttt{risc}), where the LLM successfully refactors state machines to be more compact. Conversely, optimization gains are negligible for mature, standardized interface designs like \texttt{usb} and \texttt{ethmac}. This suggests that for legacy designs with rigid behavioral specifications and minimal logic redundancy, current LLMs are limited in their ability to perform the high-level architectural transformations required for further PPA gains.
\vspace{-3pt}

\subsection{Analysis of Code Evolution} \label{Sec:Exp:Evo}
To interpret the reported PPA gains, we manually audit the code revisions produced by LLMs. We observe that the agent consistently applied three high-level architectural strategies rather than random mutations. 
\begin{itemize}
    \item \textbf{Coding Style Attempt: } The agent exploits the synthesis-friendly coding style, which is related to both toolchain and designs. By iteratively proposing functionally equivalent syntactic variations (e.g., converting blocking to non-blocking assignments, or changing loop constructs to vector operations), the framework discovers and accumulates the implicit coding patterns that the synthesis tool maps more efficiently. 
    \item \textbf{Logic Flattening and Pruning:} In computational-heavy designs like \texttt{hsm} and \texttt{audio}, the agent frequently refactored deep, serialized logic chains (e.g., nested conditional blocks) into parallelized expressions. It also identified and removed redundant intermediate pipeline registers where timing budgets allowed, directly reducing cell area and logic depth.
    \item \textbf{State Machine Optimization:} For designs consisting of control units, like \texttt{mem\_ctrl} and \texttt{risc}, the agent merges multiple disparate state flags into unified, binary-encoded state registers. It also simplifies transition logic by merging equivalent states and removing redundant states.
\end{itemize}

\subsection{Results on Commercial Toolchain} \label{Sec:Exp:DC}

\begin{table}[!t]
\caption{Evolution results with commercial toolchain} \label{Tab:DC} 
    \tabcolsep = 0.015\linewidth
\begin{small}
\begin{tabular}{@{}l|lll|lll@{}}
\toprule
            & \multicolumn{3}{c|}{Baseline}                              & \multicolumn{3}{c}{DeepSeek-v3}                                \\
Design      & Area & Delay & ADP        & Area & Delay & ADP         \\ 
      & ($\mu m^2$) & ($ns$) & (K)        & ($\mu m^2$) & ($ns$) & (K)         \\ \midrule
\texttt{audio}       & 5,846.86                     & 1.79       & 10.47          & 5,846.86                     & 1.60       & \textbf{9.35}   \\
\texttt{ethmac}      & 71,083.17                    & 3.27       & 232.44         & 70,681.54                    & 3.27       & \textbf{231.13} \\
\texttt{hsm}         & 17,845.87                    & 2.88       & 51.40          & 17,309.10                    & 2.88       & \textbf{49.85}  \\
\texttt{mem\_ctrl}   & 14,076.00                    & 3.74       & 52.64          & 13,889.57                    & 3.74       & \textbf{51.95}  \\
\texttt{nn\_engine}  & 435,505.18                   & 12.69           & 5,526.56               & 429,478.46                             & 12.69           & \textbf{5450.08}       \\
\texttt{risc}   & 158,180.45                   & 9.79       & 1,548.59       & 158,180.45                   & 9.79       & 1,548.59        \\
\texttt{sdc\_ctrl}   & 42,916.16                    & 1.63       & 69.95          & 42,916.16                    & 1.63       & 69.95           \\
\texttt{simple\_cpu} & 30,784.52                    & 8.94       & 275.21         & 30,569.32                    & 8.94       & \textbf{273.29} \\
\texttt{spi}         & 1,577.76                     & 0.50       & 0.79           & 1,577.76                     & 0.50       & 0.79            \\
\texttt{usb}         & 38,038.98                    & 2.55       & 97.00          & 37,538.50                    & 2.55       & \textbf{95.72}  \\
\texttt{vga\_enh}         & 95,469.06                    & 2.16       & 206.21         & 95,380.22                    & 2.16       & 206.02          \\ \midrule
Geo.        & 33,223.78                    & 3.16       & \textbf{105.01} & 32,969.04                    & 3.13       & \textbf{103.14}  \\
Red.        &                              &            &                & 0.77\%                       & 1.01\%     & \textbf{1.77\%} \\ \bottomrule
\end{tabular}
\end{small}
\end{table}
To validate the robustness of our proposed reference framework, we replicate the evolution experiment using a commercial toolchain. We utilize the Synopsys Design Compiler command \texttt{compile\_ultra} with maximum optimization flags enabled, specifically including \texttt{--retime} and \texttt{--timing\_high\_effort\_script}. We maintain the same experimental setup as in Section~\ref{Sec:Exp:Yosys}, employing DeepSeek-v3 as the base model, as it demonstrated the superior performance with the open-source toolchain. The results are summarized in Table~\ref{Tab:DC}.

Despite the effectiveness, the overall magnitude of improvement is smaller compared to the gains obtained with the open-source toolchain. The primary reason for the diminished gain is the aggressive optimization capability of commercial synthesis tools. In many cases, the LLM successfully produces reasonable changes on RTL code. While these changes are syntactically distinct and yield gains in open-source tools, the powerful commercial tools often reduces both the original and the refactored code to the same gate-level netlist. 

While the geometric mean reduction of 1.77\% is numerically smaller than the open-source results, it represents a significant engineering achievement given the baseline. To the best of our knowledge, we employ the state-of-the-art logic optimization and synthesis tool (i.e. \texttt{compile\_ultra}) with all high effort algorithms enabled. Notably, the framework still further evolves the \texttt{audio} design reducing its critical path delay by 10.61\%. Besides, it achieves a reduction of 536.77 $\mu m^2$ on the \texttt{hsm} design. We note that RTL evolution is one-time offline investment. Given that industrial PPA margins are typically extremely tight, an automated improvement of this magnitude on top of logic synthesis and optimization represents a valuable gain for practical chip design scenarios.

\section{Future Work}
\methodname~establishes a baseline for repo-level RTL evolution, yet our experimental results highlight several open challenges for future research. 

\begin{itemize}
    \item \textbf{From Local Optimization to Architectural Refactoring}: As observed in Section~\ref{Sec:Exp:DC}, current LLMs optimizing local coding styles overlap with the capabilities of advanced synthesis engines. To achieve better PPA reduction with commercial tools, future evolution agents must move beyond module-level rewriting and attempt high-level architectural transformations. For example, future work should leverage the \methodname~benchmark to investigate agents capable of analyzing entire data and control flows across the repository to implement global changes and explore better architectures. 
    \item \textbf{Powerful Base Models for Evolution}: The efficacy of the evolutionary search relies on the quality of the mutation operator to propose valid improvements. While Section~\ref{Sec:Exp:Evo} demonstrates that LLMs can execute valid local changes like state machine encoding or logic flattening, they typically lack the specific knowledge required to align code with the target PDK and EDA toolchains. Future work should focus on mining explicit ``optimization rules'' from high-quality evolution traces and guiding the LLM to apply these environment-specific patterns via domain-adaptive post-training or Retrieval-Augmented Generation (RAG).
\end{itemize}

\section{Conclusion} \label{Sec:Conclusion}
In this work, we formulate the task of repo-level RTL evolution, which evolves the multi-file RTL repository to meet PPA constraint or optimize a specific metric. To spport the task, we propose \methodname~, a repository-level benchmark for the iterative evolution of RTL designs. In our benchmark, we curate high-quality RTL designs across different categories and also provide a reference closed-loop framework that integrates LLM-driven mutation. Our experiments demonstrate that current LLMs can autonomously achieve area and delay reduction on the benchmark designs. Without any manual intervention, the close-loop evolution framework can reduce 10.50\% ADP using an open-source toolflow. Furthermore, we highlight existing challenges regarding optimization strategies and base model capabilities. \methodname~establishes a foundation for future research into autonomous hardware engineering, guiding the transition of LLMs from code generators to design experts in practical applications.

\balance


\newpage

\balance
\bibliographystyle{unsrt}
\bibliography{reference}

\end{document}